\documentclass[10pt]{article}
\usepackage{graphicx}

\def\Title#1{\begin{center} {\Large #1 } \end{center}}
\def\Author#1{\begin{center}{ \sc #1} \end{center}}
\def\Address#1{\begin{center}{ \it #1} \end{center}}

\newcommand\pubblock{\rightline{\begin{tabular}{l} Proceedings of the Second Annual LHCP\\ \pubnumber\\
         \pubdate  \end{tabular}}}

\newenvironment{Abstract}{\begin{quotation} \begin{center} 
             \large ABSTRACT \end{center}\bigskip 
      \begin{center}\begin{large}}{\end{large}\end{center} \end{quotation}}

\newenvironment{Presented}{\begin{quotation} \begin{center} 
             PRESENTED AT\end{center}\bigskip 
      \begin{center}\begin{large}}{\end{large}\end{center} \end{quotation}}

\def\beq{\begin{equation}}
\def\eeq#1{\label{#1}\end{equation}}
\def\eeqn{\end{equation}}

\def\beqa{\begin{eqnarray}}
\def\eeqa#1{\label{#1}\end{eqnarray}}
\def\eeqan{\end{eqnarray}}

\let\bar=\overbar

\def\Dslash{\not{\hbox{\kern-4pt $D$}}}
\def\dslash{\not{\hbox{\kern-2pt $\del$}}}

\def\msb{{\bar{\ssstyle M \kern -1pt S}}}

\usepackage{color}

\newcommand\pubnumber{ CMS CR-2014/198 }

\newcommand\pubdate{\today}

\def\affiliation{
On behalf of the CMS Collaboration, \\
Department of Physics \\
Brown University, Providence, R.I. 02906, U.S.A }

\begin{document}

\large
\begin{titlepage}
\pubblock

\vfill
\Title{Performance of Muon-Based Triggers at the CMS High Level Trigger}
\vfill

\Author{ Juliette Alimena }
\Address{\affiliation}
\vfill
\begin{Abstract}

The trigger systems of the CERN LHC detectors play a crucial role in determining the physics capabilities of the experiments. 
A reduction of several orders of magnitude of the event rate is needed to reach values compatible with the detector readout, offline storage and analysis capabilities. 
The CMS experiment has been designed with a two-level trigger system: the Level 1 (L1) Trigger, implemented on custom-designed electronics, and the High Level Trigger (HLT), 
a streamlined version of the CMS reconstruction and analysis software running on a computer farm. 
Here we will present the design and performance of the main muon triggers used during the Run I data taking. 
We will show how these triggers contributed to the 2012 physics results. 
We will then present the improvements foreseen to meet the challenges of the Run II data taking. 
We will discuss the improvements being made at L1, and at various stages in the HLT reconstruction, 
ranging from the local drift tube and cathode strip chamber reconstruction, to L2 muon tracks, to the final L3 muons.

\end{Abstract}
\vfill

\begin{Presented}
The Second Annual Conference\\
 on Large Hadron Collider Physics \\
Columbia University, New York, U.S.A \\ 
June 2-7, 2014
\end{Presented}
\vfill
\end{titlepage}
\def\thefootnote{\fnsymbol{footnote}}
\setcounter{footnote}{0}
%

\normalsize 


\section{Introduction}

The CMS muon system consists of Drift Tubes (DTs), which cover $|\eta|<1.2$, 
Resistive Plate Chambers (RPCs), which cover $|\eta|<1.6$, and Cathode Strip Chambers (CSCs), which cover $0.9<|\eta|<2.4$.
All three muon subdetectors are used in the CMS trigger. The CMS trigger consists of the Level 1 (L1) Trigger, which is hardware based and uses the muon detectors only, 
and the High Level Trigger (HLT), which is software based and uses the muon, calorimter, and tracker detectors.

At the HLT, there are two main steps in the muon reconstruction. The first is the Level 2 (L2) muon reconstruction, 
which builds tracks in the muon system, and the second is the Level 3 (L3) muon reconstruction, 
which builds full tracks from the L2 muon tracks and the tracker information.

The L2 reconstruction starts by building seeds from patterns of DT and CSC segments. Then, the reconstruction of a track is 
started from these seeds, using measurements from all of the muon chambers. Finally, filters on the track quality, $\eta$, and $p_{T}$ of the L2 muons are applied in 
order to reduce the rate.

The L3 reconstuction builds full muon tracks from the muon system and the tracker. The L3 reconstruction starts by building a seed for the tracker 
reconstruction, starting from the L2 information. Then, the tracker track is reconstructed. Finally, the tracker track is matched to the 
L2 muon. Different seeding algorithms are tried in the L3 cascade algorithm (see below). Furthermore, filters on the track quality, $\eta$, and $p_{T}$ of the L3 muons
 are applied in order to reduce the rate.

Muon triggers require one or more candidates (Single Muon Triggers) or two or more candidates (Double Muon Triggers) and use isolation,
 good track quality, good vertices, and other requirements in order to select muons. Isolation can be measured by searching for tracks 
and calorimeter deposits in a cone around the L3 muon.

\section{Run I Single Muon Triggers Performance}

\begin{figure}[htb]
\centering
\includegraphics[height=1.7in]{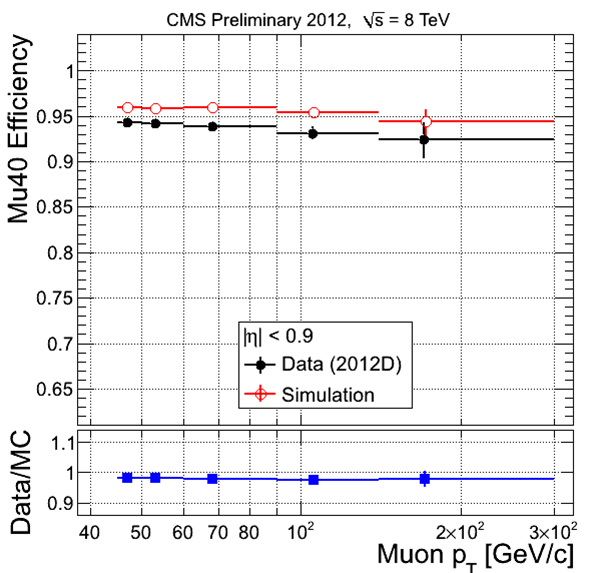}
\includegraphics[height=1.7in]{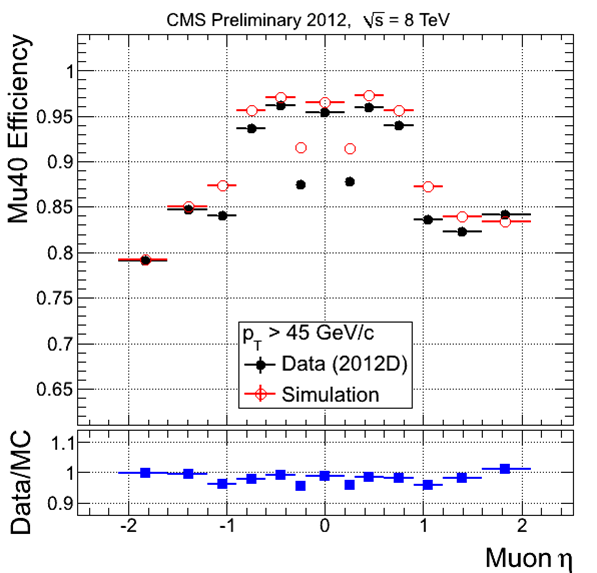}
\includegraphics[height=1.7in]{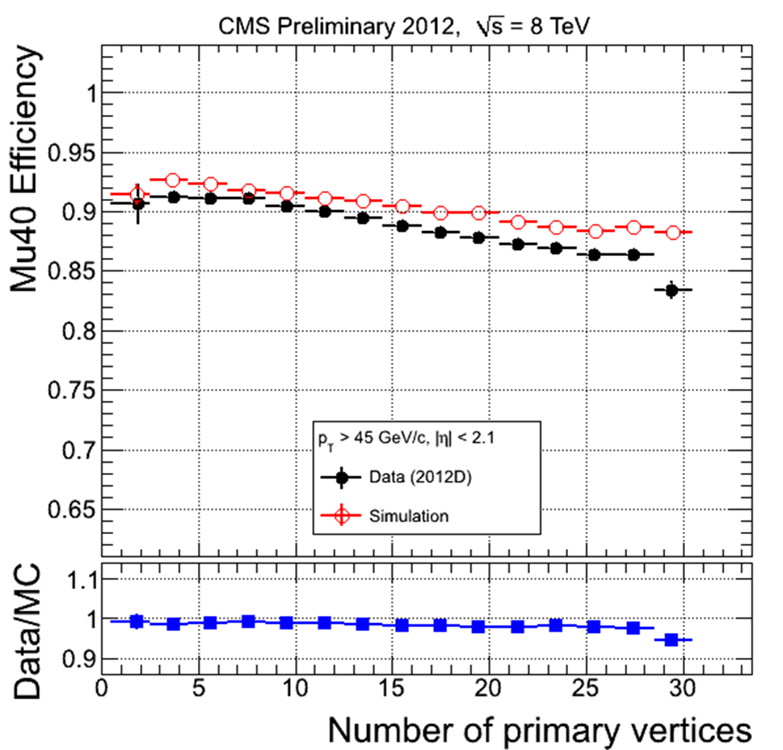}
\caption{Efficiency of HLT\_Mu40 in 2012, for data and Monte Carlo (MC) simulation. The efficiency is computed with the Z resonance tag and probe method. 
The efficiency is computed with respect to muons that pass the tight identification criteria described in \cite{CMS-DP-2013-009} and have 
a $p_T$ of at least 45 GeV. In addition, for the muon $p_T$ plot, the muon is required to be within $|\eta|<0.9$, 
and for the number of primary vertices plot, the muon is required to be within $|\eta|<2.1$.}
\label{fig:figure1}
\end{figure}

Figure~\ref{fig:figure1} shows the efficiency of HLT\_Mu40, the unprescaled, non-isolated muon HLT path with the lowest $p_T$ threshold used in the 2012 run. This HLT path 
requires at least one L3 muon with a $p_T$ of 40 GeV.

\section{Improvements for Run II: L3 Muon Triggers}

In preparation for Run II, a number of improvements were made to the muon triggers. In particular, improvements were made to the L3 muon trigger algorithms.
This was done in order to recover the efficiency loss for L3 muon triggers at high pileup (PU). The efficiency at high PU was improved by implementing 
changes in the so-called L3 cascade algorithm.

\begin{figure}[htb]
\centering
\includegraphics[height=2in]{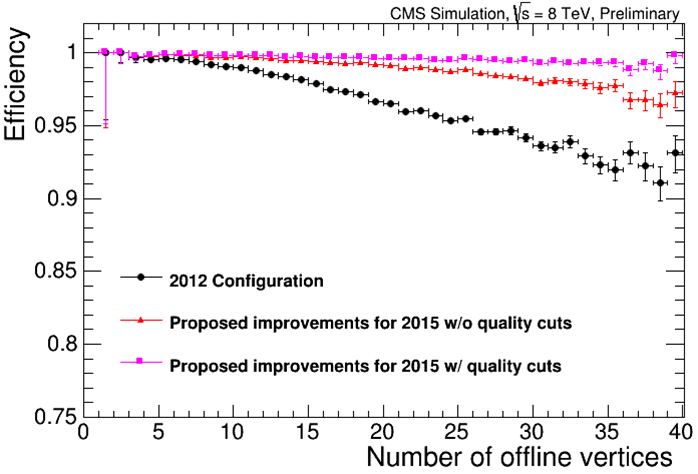}
\caption{The efficiency for the L3 cascade step, before and after the cascade algorithm improvements, as a function of the number of offline vertices. 
The denominator is the number of generated 
muons matched with a L2 track and passing the HLT quality cuts. The numerator is the number of generated muons as above, that also match a L3 track passing the final L3 muon filter 
quality cuts and whose tracker track component has $>75\%$ hits shared with a simulated muon track.}
\label{fig:figure2}
\end{figure}

In essence, three algorithms are tried in ``cascade'', from the fastest 
to the most CPU-consuming one, until a L3 track is found. 
In the default version of the L3 cascade algorithm, the sequence stops as soon as a L3 track was built. However, it is possible that an algorithm of the cascade can build a track that fails 
the final set of cuts applied at the end of the L3 reconstruction, 
but the next algorithms might do better. Therefore, the L3 cascade algorithm was improved by filtering on quality cuts in each algorithm of the cascade, before the 
tracker track is matched to the L2 muon. 
Furthermore, the L3 reconstruction configuration was also improved in order to increase the efficiency at high PU independently from the requirement of quality cuts within the L3 cascade.
For example, the criteria to define compatibility of hits with a given trajectory at the track building stage was changed from the one used in 2012 to one more robust against PU. 
Figure~\ref{fig:figure2} shows the efficiency of the L3 step, before and after the cascade algorithm improvements, using W$\rightarrow\mu\nu$ simulated events.

The trigger rates do increase slightly, when moving from the 2012 configuration to the proposed improvements for 2015 with the quality cuts. The rate of HLT\_IsoMu24 increases 
by 4.3\%, and the rate of HLT\_Mu40 increases by 6.8\%. This rate increase is expected, as the efficiency is improving, and furthermore, it was deemed 
acceptable for data taking in 2015.

\section{Improvements for Run II: Muon Triggers Isolation}

Another major improvement for 2015 data taking was the isolation improvement in muon triggers. We will first discuss isolation in single muon triggers, and then
we will discuss isolation in double muon triggers.

\subsection{Single Muon Isolation}

\begin{figure}[htb]
\centering
\includegraphics[height=2in]{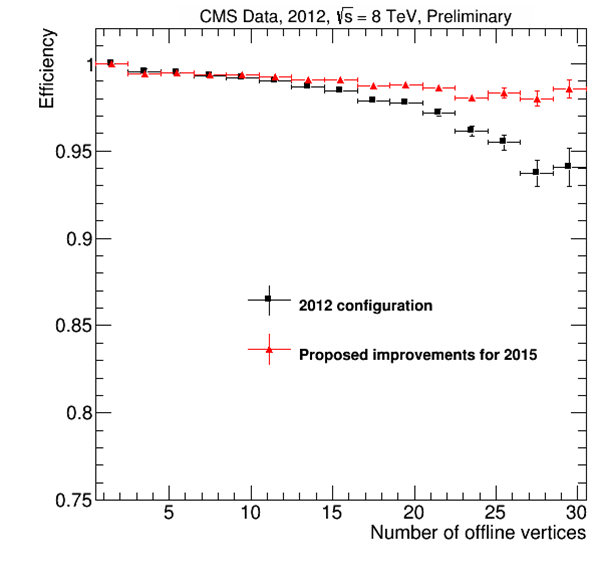}
\includegraphics[height=2in]{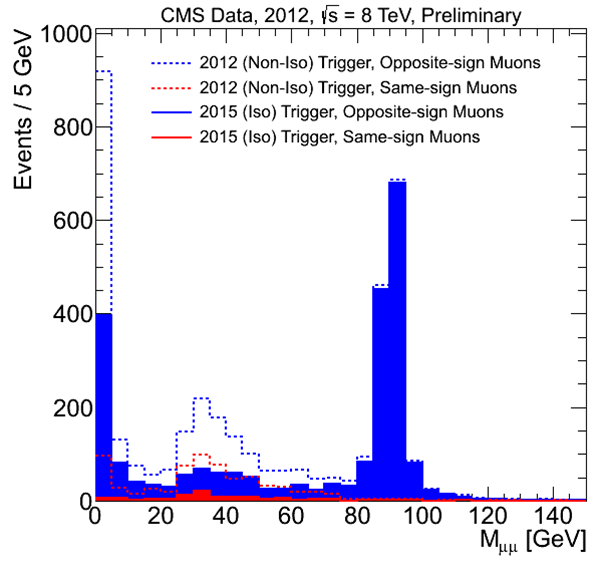}
\caption{(Left) The efficiency in single muon isolated triggers, for the 2012 and proposed 2015 configurations, as a function of the number of offline vertices. 
The efficiency is computed with respect to reconstructed muons that pass the tight muon identification and loose muon isolation as described in \cite{CMS-DP-2013-009}. 
Futhermore, the muon must have at least 25 GeV in $p_T$, $|\eta|<2.1$, and be matched to HLT\_IsoMu24\_eta2p1.
(Right) The invariant mass distribution for muons passing the isolated and non-isolated versions of HLT\_Mu17\_Mu8.}
\label{fig:figure3}
\end{figure}

The single muon trigger isolation was improved in order to recover the efficiency loss at high PU and to reduce the CPU time. This was done by optimizing the PU 
mitigation and tracking configurations. Figure~\ref{fig:figure3} (left) shows the efficiecny as a function of number of vertices, for the 2012 configuration and for the 
proposed improvements for 2015. The overall efficiency improves, especially at high PU, while the rate reduction which comes from applying isolation is 
constant between the two configurations. Furthermore, the new tracking configuration used in the isolation reduces the CPU time with respect to the one used in 2012.
The mean HLT path time for HLT\_IsoMu24\_eta2p1, when running on an event skim selected by requiring the non isolated version of the trigger, 
moves from 836 ms in the 2012 configuration to 83.9 ms in the 2015 proposal.

\subsection{Double Muon Isolation}

Loose tracker isolation was introdued in the double muon triggers in order to reduce the rate of these triggers. The efficiency of the triggers remained at
greater than 99\%, using the definitions of ``loose'' and ``tight'' muon identification described in \cite{stop_search} and \cite{Higgs_measurement}, while the rate was reduced by 44\%. 
Figure~\ref{fig:figure3} (right) shows the invariant mass distribution for muons passing isolated and 
non-isolated double muon triggers. The efficiency in the Z peak is maintained with the introduction of isolation, but the rate is reduced for low mass events.

\section{Conclusions}
The performance of the CMS muon triggers in 2012 was presented, along with the possible improvements for 2015.


\begin{thebibliography}{99}


  
\bibitem{CMS-DP-2013-009} 
  S.~Chatrchyan {\it et al.}  [CMS Collaboration],
  CERN Document Server (2013), CMS-DP-2013-009,
  [https://cds.cern.ch/record/1536406?ln=en].

\bibitem{stop_search} 
  S.~Chatrchyan {\it et al.}  [CMS Collaboration],
  EPJ\ C {\bf 73}, (2013)
  [DOI:10.1140/epjc/s10052-013-2677-2].

\bibitem{Higgs_measurement} 
  S.~Chatrchyan {\it et al.}  [CMS Collaboration],
  Phys.\ Rev.\ D {\bf 89}, 092007 (2014)
  [DOI:10.1103/PhysRevD.89.092007].


\end{thebibliography}
\end{document}